\documentclass[11pt,a4paper]{scrartcl}

\usepackage{CLICdp}
\usepackage{float}
\usepackage{amsmath}

\usepackage{CLICdp_definitions}

\RequirePackage{heppennames2}

\newcommand{\MeV}{\ensuremath{\text{MeV}}\xspace}
\newcommand{\GeV}{\ensuremath{\text{GeV}}\xspace}
\newcommand{\TeV}{\ensuremath{\text{TeV}}\xspace}
\newcommand{\tabt}[1]{\multicolumn{1}{c}{#1}}
\newcommand{\tabtt}[1]{\multicolumn{2}{c}{#1}}
\newcommand{\mH }{\ensuremath{m_{\PH}}\xspace}
\newcommand{\GH }{\ensuremath{\Gamma_{\PH}}\xspace}
\newcommand{\BR}{\ensuremath{BR}\xspace}

\newcommand{\gHZZ}{\ensuremath{g_{\PH\PZ\PZ}}\xspace}
\newcommand{\gHWW}{\ensuremath{g_{\PH\PW\PW}}\xspace}
\newcommand{\gHtt}{\ensuremath{g_{\PH\PQt\PQt}}\xspace}
\newcommand{\gHbb}{\ensuremath{g_{\PH\PQb\PQb}}\xspace}
\newcommand{\gHcc}{\ensuremath{g_{\PH\PQc\PQc}}\xspace}
\newcommand{\gHTauTau}{\ensuremath{g_{\PH\PGt\PGt}}\xspace}
\newcommand{\gHMuMu}{\ensuremath{g_{\PH\PGm\PGm}}\xspace}

\title{CLIC Higgs coupling prospects with a longer first energy stage}

\clicdpnote{2020}{001}  

\date{\today}

\addauthor{A.~Robson}{\institute{1}}
\addauthor{P.~Roloff}{\institute{2}}
\addauthor{J.~de Blas}{\institute{3}}
\addinstitute{1}{University of Glasgow, Glasgow, United Kingdom}
\addinstitute{2}{CERN, Geneva, Switzerland}
\addinstitute{3}{Institute for Particle Physics Phenomenology, Durham University, Durham, United Kingdom}

\abstract{One of the most attractive features of a linear collider is the ability to
  extend its energy reach in stages, and to adapt the running plan flexibly
  in terms of maximum centre-of-mass energy and time spent at each stage.  
  The baseline luminosity staging scenario for CLIC is well-established, and has been 
  used to obtain sensitivity projections for Standard Model measurements
  and Beyond Standard Model scenarios.
  Here, as an exercise to illustrate what could be obtained from an alternative
  running scenario, Higgs coupling sensitivities are presented for the case where
  more data is collected at the initial stage
  $\roots=380\,\GeV$, before proceeding to the higher energy stages of $\roots=1.5$ and $3\,\TeV$. 
  This could be achieved through running for longer, or operating
  the collider at an increased repetition rate of 100\,Hz at the initial stage, or a combination of both.}

\graphicspath{ {./logos/}{./figures/} }

\begin{document}

\titlepage

\section{Introduction}

The Compact Linear Collider, CLIC, offers high-energy \epem\ collisions up to centre-of-mass energies of 3\,TeV \cite{clic-study}.
A rich programme of Higgs and top-quark physics is uniquely provided by the initial energy
stage around $\roots=380\,\GeV$; this is supplemented at the higher-energy stages
by increased precision in Higgs and top-quark physics, and further reach to Beyond Standard Model (BSM)
effects \cite{Abramowicz:2016zbo,Abramowicz:2018rjq,deBlas:2018mhx}. 

One of the most attractive features of a linear collider is the ability to
extend the energy reach in stages.  Correspondingly, this provides a high degree of
flexibility in adapting the programme in terms both of maximum centre-of-mass energy and running times at each stage;
this could be in response to physics results that suggest reoptimisation;
to technical developments in CLIC or other accelerator technologies
that suggest updating the global accelerator strategy; or to other considerations
such as the availability of funding, which could change the schedule.

In the context of the European Strategy for Particle Physics, while
the CLIC proposal presents a sequence of energy upgrades from $\roots=380\,\GeV$ (CLIC$_{380}$)
to 1.5\,TeV (CLIC$_{1500}$) and 3\,TeV (CLIC$_{3000}$), this flexibility means that it would also be possible to take a different
route after the initial stage, for example:
\begin{equation*}
\begin{aligned}
  \text{CLIC}_{380}  &+  \text{CLIC}_{1500} + \text{CLIC}_{3000}  \\
  \text{CLIC}_{380}  &+  \text{CLIC}_{1500} + \text{FCC}_{hh}  \\
  \text{CLIC}_{380}  &+  \text{FCC}_{hh}   \\
  \text{CLIC}_{380}  &+  \text{muon collider} \\  
  \text{CLIC}_{380}  &+  \text{wakefield acceleration} \\  
  \text{CLIC}_{380}  &+  \text{dielectric-based acceleration.}  
\end{aligned}
\end{equation*}

Starting with CLIC$_{380}$ gives the option of reviewing the physics and technology landscape
every few years, and choosing the best next step at that time, so the initial choice becomes:
\begin{equation*}
\begin{aligned}
  \text{CLIC}_{380}  &+  \text{best next step. \;\;\;\;\;\;\;} 
\end{aligned}
\end{equation*}

To explore the options afforded by a flexible running scenario, as an exercise we examine here the extra sensitivity to
Higgs couplings that would be obtained by collecting more integrated luminosity at $\roots=380\,\GeV$.  
This could arise through changed accelerator parameters -- for example running at the initial stage
with a repetition rate of 100\,Hz, which is double the baseline repetition rate --
or through running for a longer time, or a combination of both.

\section{Staging}
\label{sec:staging}

CLIC's luminosity baseline was presented in \cite{Robson:2018zje,CLICsummaryYR}, where corresponding 
Higgs coupling sensitivities were given.
The luminosity baseline assumes $1\,\abinv$ of integrated luminosity collected at the initial stage,
$\sqrt{s}=380\,\GeV$, followed by $2.5\,\abinv$ at $\sqrt{s}=1.5\,\TeV$ and
$5\,\abinv$ at $\sqrt{s}=3\,\TeV$.  The projected timescale includes a three-year
ramp-up to reach the nominal instantaneous luminosity for the first energy stage,
and two-year ramp-ups at the second and third stages.  

For a comprehensive mapping of the Higgs\,(125\,GeV) sector in \epem collisions, some data must
be taken close to the threshold for Higgs production, where the Higgs-strahlung
process dominates.  This allows a precise measurement of the total Higgs production
cross-section through the so-called `recoil' method, which is needed in order to
extract Higgs couplings in a model-independent way from measured cross-sections times branching ratios.

The CLIC physics priority is to move to higher centre-of-mass energies quickly,
to take advantage of the unique capability of CLIC to provide multi-TeV \epem 
collisions.  The higher-energy stages open production channels not accessible at
the initial energy stage and provide enhanced sensitivity to BSM scenarios.
The CLIC baseline luminosity scenario therefore moves to $\sqrt{s}=1.5\,\TeV$ after
8 years of running at the initial energy stage.

However, it is useful to see how sensitivities would be affected by taking more data 
at the initial stage, for example to adapt to the available funding profile.

Owing to the initial ramp-up in instantaneous luminosity, integrated luminosity is
accumulated more quickly later in the run, after the nominal instantaneous luminosity has been reached.
Therefore, by increasing the initial stage
from 8 years to 13 years, the integrated luminosity is doubled.  Furthermore, as
discussed in \cite{SteinarPostGranada}, it would be possible to operate CLIC at $380\,\GeV$
at double the repetition
rate -- 100\,Hz instead of 50\,Hz -- with only modest increase in
cost (around the 5\% level) and power (from around 170\,MW to around 220\,MW).

Taking both of these enhancements into account, as an exercise we consider sensitivities resulting
from $4\,\abinv$ collected at the initial stage, rather than the baseline $1\,\abinv$.
As in the baseline scenario, equal amounts of --80\% and +80\% polarisation running are 
foreseen throughout the initial energy stage.

The two examples of 
(a) the CLIC baseline of $1\,\abinv$ collected at $380\,\GeV$ plus $2.5\,\abinv$ collected at $1.5\,\TeV$
(presented in \cite{Robson:2018zje} and \cite{CLICsummaryYR}), and
(b) $4\,\abinv$ collected at $380\,\GeV$ (presented here), provide realistic scenarios
that can usefully be compared 
with other proposed \epem collider options that are limited in centre-of-mass energy.

\section{Higgs couplings}
\label{sec:higgs}

The total Higgs production cross section measurement $\sigma(\PZ\PH)$, using only the
system that recoils against the produced Higgs boson and without examining the Higgs
decay products, is a unique feature of lepton colliders.
It dominates the model-independent determination of the ZH coupling, $\gHZZ$, and is
only possible at the initial energy stage. 
In turn this propagates into the extraction of all the other Higgs couplings.
Accumulating more data at $\roots=380\,\GeV$ therefore contributes to improved
precision on the other Higgs couplings.

\subsection{Summary of Higgs observables}

Extensive studies of the CLIC sensitivities to Higgs couplings have been reported
previously in \cite{Abramowicz:2016zbo}, where details of the analyses and
the extraction of Higgs observables through combined
fitting can be found{\footnote{Note that earlier studies assumed an energy staging of $\sqrt{s}=350\,\GeV$, 1.4\,\TeV, and 3\,\TeV; those energy stages are used for the results presented here, but with results scaled to the updated integrated luminosities.}}.
Sensitivities obtained assuming the current CLIC luminosity baseline can be found in \cite{Robson:2018zje}.

Here, the precisions of the individual Higgs sector measurements are given for an
increased luminosity of $4\,\abinv$ at the initial energy stage, while
the luminosities of the 1.4 (1.5) and 3\,TeV stages are unchanged at 
2.5 and 5.0\,\abinv, respectively.  This serves to illustrate what could be
obtained from the different stages of an alternative running scenario.

Precisions on the Higgs observables are given in \autoref{tab:GlobalFit:Input350} for
the first energy stage, and in \autoref{tab:GlobalFit:Input143} for the two
higher-energy stages.  These individual results assume unpolarised beams.

Measurement of the cross section for double-Higgs production at 1.4 and 3\,TeV 
gives sensitivity to the Higgs self-coupling $\lambda$.
This is unchanged from that reported in \cite{Robson:2018zje,Roloff:2019crr}, with an ultimate precision on 
 $\lambda$ of $[-7\%,+11\%]$.

The recoil mass analysis from $\Pep\Pem\to\PZ\PH$
events can be used to search for BSM decay modes of the Higgs boson into 
`invisible' final states.  Scaling the result from \cite{Abramowicz:2016zbo}
to 4\,ab$^{-1}$ at $\roots=350\,\GeV$ 
gives an upper limit
on the invisible Higgs branching ratio of $BR$($\PH\to$ invis.) $<$ 0.34\%  
at 90\% C.L. in the modified frequentist approach. 

\begin{table*}[htp]\centering
  \begin{tabular}{lllcc}\toprule
                        &                                                           &                              & \tabt{Statistical precision} & \\ \cmidrule(l){4-4}
        \tabt{Channel}  & \tabt{Measurement}                                        & \tabt{Observable}            & $350\,\GeV$ & Reference \\ 
                        &                                                           &                              & $4\,\abinv$ & \cite{Abramowicz:2016zbo} \\ \midrule
    $\PZ\PH$            & Recoil mass distribution                                  & $\mH$                        & $39\,\MeV$ & \cite{Abramowicz:2016zbo} \\
    $\PZ\PH$            & $\sigma(\PZ\PH)\times \BR(\PH\to\text{invisible})$        & $\Gamma_\text{inv}$           & $0.2\,\%$ & \cite{Abramowicz:2016zbo} \\ \midrule
    $\PZ\PH$            & $\sigma(\PZ\PH)\times \BR(\PZ\to\Plp\Plm)$                & $\gHZZ^{2}$                   & $1.3\,\%$ & \cite{Abramowicz:2016zbo} \\
    $\PZ\PH$            & $\sigma(\PZ\PH)\times \BR(\PZ\to\PQq\PAQq)$               & $\gHZZ^{2}$                   & $0.6\,\%$ & \cite{Abramowicz:2016zbo} \\
    $\PZ\PH$            & $\sigma(\PZ\PH)\times \BR(\PH\to\PQb\PAQb)$               & $\gHZZ^{2}\gHbb^{2}/\GH$       & $0.30\,\%$ & \cite{Abramowicz:2016zbo} \\
    $\PZ\PH$            & $\sigma(\PZ\PH)\times \BR(\PH\to\PQc\PAQc)$               & $\gHZZ^{2}\gHcc^2/\GH$        & $5\,\%$ & \cite{Abramowicz:2016zbo} \\
    $\PZ\PH$            & $\sigma(\PZ\PH)\times \BR(\PH\to\Pg\Pg)$                  &                              & $2.2\,\%$ & \cite{Abramowicz:2016zbo} \\
    $\PZ\PH$            & $\sigma(\PZ\PH)\times \BR(\PH\to\tptm)$                   & $\gHZZ^{2}\gHTauTau^{2}/\GH$  & $2.2\,\%$ & \cite{Abramowicz:2016zbo} \\
    $\PZ\PH$            & $\sigma(\PZ\PH)\times \BR(\PH\to\PW\PW^*)$                & $\gHZZ^{2}\gHWW^{2}/\GH$      & $1.8\,\%$ & \cite{Abramowicz:2016zbo} \\
    $\PH\PGne\PAGne$    & $\sigma(\PH\PGne\PAGne)\times \BR(\PH\to\PQb\PAQb)$       & $\gHWW^{2}\gHbb^{2}/\GH$      & $0.7\,\%$ & \cite{Abramowicz:2016zbo} \\
    $\PH\PGne\PAGne$    & $\sigma(\PH\PGne\PAGne)\times \BR(\PH\to\PQc\PAQc)$       & $\gHWW^{2}\gHcc^{2}/\GH$      & $9\,\%$ & \cite{Abramowicz:2016zbo} \\
    $\PH\PGne\PAGne$    & $\sigma(\PH\PGne\PAGne)\times \BR(\PH\to\Pg\Pg)$          &                             & $3.5\,\%$ & \cite{Abramowicz:2016zbo} \\    
    \bottomrule
  \end{tabular}
    \caption{Summary of the precisions obtainable for the Higgs
      observables in the first stage of CLIC for an increased integrated
      luminosity of $4\,\abinv$ at $\roots=350\,\GeV$, assuming
      unpolarised beams, to illustrate the scenario where more data is taken
      at the inital energy stage than in the CLIC baseline. 
      For the branching ratios, the measurement
      precision refers to the expected statistical uncertainty on the
      product of the relevant cross section and branching ratio; this
      is equivalent to the expected statistical uncertainty of the
      product of couplings divided by $\Gamma_{\PH}$ as indicated in
      the third column. \label{tab:GlobalFit:Input350}}
\end{table*}

\begin{table*}[htp]\centering
    \begin{tabular}{lllccc}\toprule
                        &                                                            &                             & \tabtt{Statistical precision} & \\ \cmidrule(l){4-5}
        \tabt{Channel}  & \tabt{Measurement}                                         & \tabt{Observable}           & $1.4\,\TeV$        & $3\,\TeV$ & Reference \\ 
                        &                                                            &                             & $2.5\,\abinv$      & $5.0\,\abinv$ & \\ \midrule
   $\PH\PGne\PAGne$     & $\PH\to\PQb\PAQb$ mass distribution                        & $\mH$                       & $36\,\MeV$         & $28\,\MeV$ & \cite{Abramowicz:2016zbo} \\ \midrule
   $\PZ\PH$             & $\sigma(\PZ\PH)\times \BR(\PH\to\PQb\PAQb)$                & $\gHZZ^{2}\gHbb^{2}/\GH$     & $2.6\,\%^{\dagger}$   & $4.3\,\%^{\dagger \ddagger}$ & \cite{Ellis:2017kfi} \\
   $\PH\PGne\PAGne$     & $\sigma(\PH\PGne\PAGne)\times \BR(\PH\to\PQb\PAQb)$        & $\gHWW^{2}\gHbb^{2}/\GH$     & $0.3\,\%$          & $0.2\,\%$ & \cite{Abramowicz:2016zbo} \\
   $\PH\PGne\PAGne$     & $\sigma(\PH\PGne\PAGne)\times \BR(\PH\to\PQc\PAQc)$        & $\gHWW^{2}\gHcc^{2}/\GH$     & $4.7\,\%$          & $4.4\,\%$ & \cite{Abramowicz:2016zbo} \\
   $\PH\PGne\PAGne$     & $\sigma(\PH\PGne\PAGne)\times \BR(\PH\to\Pg\Pg)$           &                             & $3.9\,\%$          & $2.7\,\%$ & \cite{Abramowicz:2016zbo} \\
   $\PH\PGne\PAGne$     & $\sigma(\PH\PGne\PAGne)\times \BR(\PH\to\tptm)$            & $\gHWW^{2}\gHTauTau^{2}/\GH$  & $3.3\,\%$         & $2.8\,\%$ & \cite{Abramowicz:2016zbo} \\
   $\PH\PGne\PAGne$     & $\sigma(\PH\PGne\PAGne)\times \BR(\PH\to\mpmm)$            & $\gHWW^{2}\gHMuMu^{2}/\GH$    & $29\,\%$          & $16\,\%$ & \cite{Abramowicz:2016zbo} \\
   $\PH\PGne\PAGne$     & $\sigma(\PH\PGne\PAGne)\times \BR(\PH\to\upgamma\upgamma)$ &                             & $12\,\%$          & $6\,\%^*$ & \cite{Abramowicz:2016zbo} \\
   $\PH\PGne\PAGne$     & $\sigma(\PH\PGne\PAGne)\times \BR(\PH\to\PZ\upgamma)$      &                             & $33\,\%$          & $19\,\%^*$ & \cite{Abramowicz:2016zbo} \\
   $\PH\PGne\PAGne$     & $\sigma(\PH\PGne\PAGne)\times \BR(\PH\to\PW\PW^*)$         & $\gHWW^{4}/\GH$              & $0.8\,\%$         & $0.4\,\%^*$ & \cite{Abramowicz:2016zbo} \\
   $\PH\PGne\PAGne$     & $\sigma(\PH\PGne\PAGne)\times \BR(\PH\to\PZ\PZ^*)$         & $\gHWW^{2}\gHZZ^{2}/\GH$      & $4.3\,\%$         & $2.5\,\%^*$ & \cite{Abramowicz:2016zbo} \\
   $\PH\epem$           & $\sigma(\PH\epem)\times \BR(\PH\to\PQb\PAQb)$              & $\gHZZ^{2}\gHbb^{2}/\GH$      & $1.4\,\%$         & $1.5\,\%^*$ & \cite{Abramowicz:2016zbo} \\ \midrule
   $\PQt\PAQt\PH$       & $\sigma(\PQt\PAQt\PH)\times \BR(\PH\to\PQb\PAQb)$          & $\gHtt^{2}\gHbb^{2}/\GH$      & $5.7\,\%$         & $-$ & \cite{Abramowicz:2018rjq} \\
  \bottomrule
  \end{tabular}
    \caption{Summary of the precisions obtainable for the Higgs
    observables in the higher-energy CLIC stages for integrated
    luminosities of $2.5\,\abinv$ at $\roots=1.4\,\TeV$, and
    $5.0\,\abinv$ at $\roots=3\,\TeV$. In both cases unpolarised beams
    have been assumed.  These are the same sensitivities given in \cite{Robson:2018zje}.
    For $\gHtt$, the $3\,\TeV$ case has not yet been studied. 
    Numbers marked with $*$ are extrapolated from $\roots=1.4\,\TeV$
    to $\roots=3\,\TeV$ while $\dagger$ indicates projections based on fast simulations.
    For the branching ratios, the measurement precision refers to the expected
    statistical uncertainty on the product of the relevant cross
    section and branching ratio; this is equivalent to the expected
    statistical uncertainty of the product of couplings divided by
    $\Gamma_{\PH}$, as indicated in the third column. \label{tab:GlobalFit:Input143}
    $^{\ddagger}$ The value for $\sigma(\PZ\PH)\times \BR(\textrm{all hadronic})$ at 3\,TeV has recently been confirmed as 4\% in a full-simulation study \cite{Leogrande:2019dzm}.}
\end{table*}

\subsection{Combined fits}

Precisions on the Higgs couplings and width extracted from a model-independent
global fit, described in \cite{Abramowicz:2016zbo}, are given
in \autoref{tab:MIResultsPolarised8020}. 
The fit assumes the baseline scenario for beam polarisation,
but an increased integrated luminosity at the initial stage for illustration. 
The increase in cross-section from having a predominantly negatively-polarised
electron beam is taken into account by multiplying the event rates for all
$\PW\PW$-fusion measurements by a factor of 1.48, corresponding to a factor of
1.8 for $80\,\%$ of the statistics and 0.2 for the remaining $20\,\%$. 
This approach is
conservative because it assumes that all backgrounds, including those from
$s$-channel processes, which do not receive the same polarisation enhancement,
scale by the same amount.

Each energy stage contributes significantly to the Higgs programme: 
the initial stage provides $\gHZZ$ and couplings to most fermions and
bosons, while the higher-energy stages improve them and add the top-quark,
muon, and photon couplings.  The precision on $\gHZZ$ is determined by the
statistics at the initial stage.

Precisions extracted from a model-dependent
global fit, also described in \cite{Abramowicz:2016zbo}, are given in \autoref{tab:MDResultsPolarised8020}.
This fit also assumes the baseline scenario for beam polarisation,
but an increased integrated luminosity at the initial stage for illustration. 

\begin{figure}
\begin{center}
\includegraphics[width=12cm]{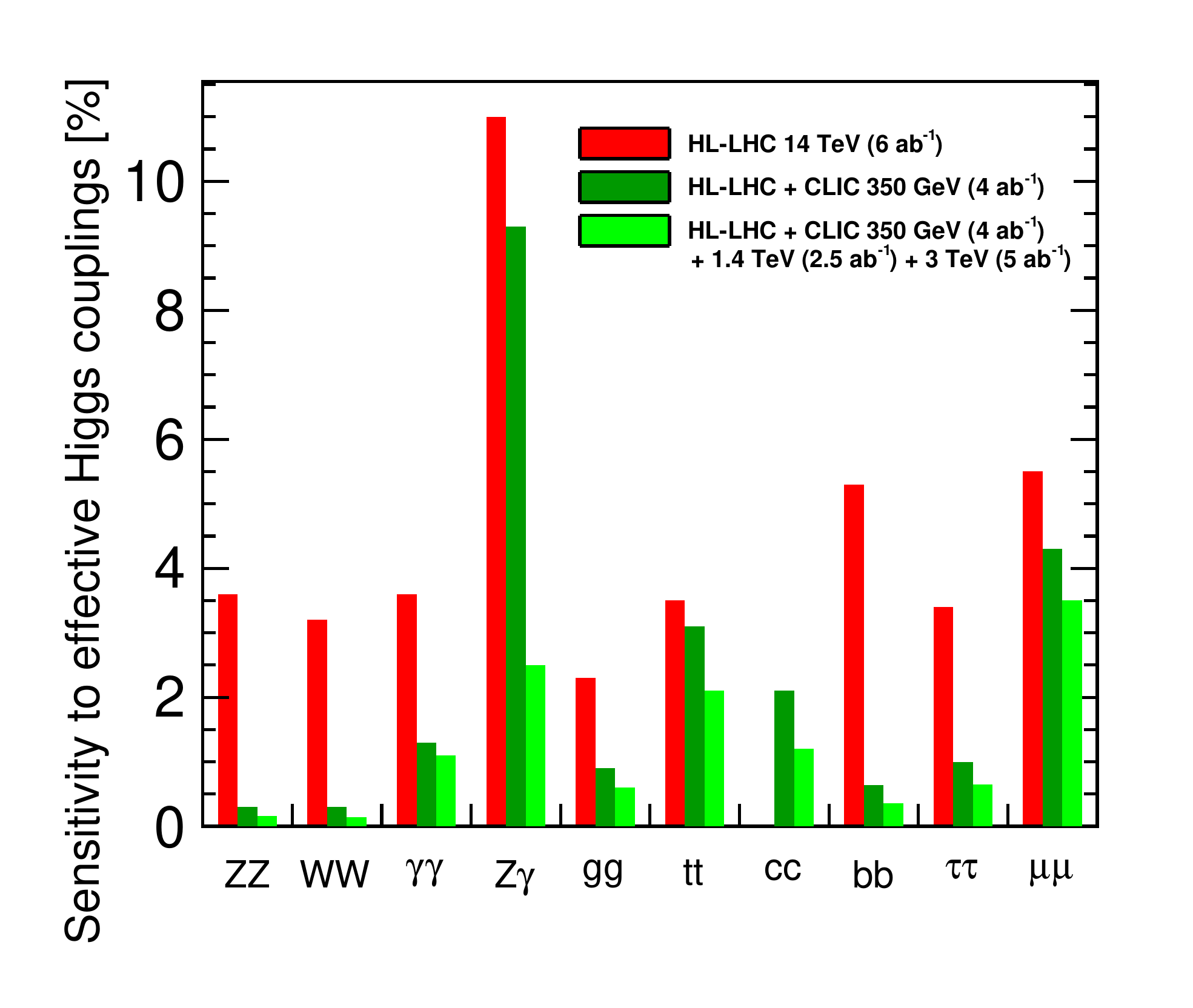}
\caption{Comparison of the SMEFT$_{\textrm{ND}}$ projections for HL-LHC and HL-LHC + CLIC. \label{fig:hllhcComparison}}
\end{center}
\end{figure}

A global EFT fit has been carried out in \cite{deBlas:2019rxi} for the purposes of comparing
future collider projects, and is extensively described there.
The corresponding projections for the illustrative increased integrated luminosity at the initial
CLIC stage, combined with the projected HL-LHC sensitivities, are given in \autoref{tab:eft-global} and \autoref{fig:hllhcComparison} 
for the model SMEFT$_{\textrm{ND}}$, which does not assume flavour universality. 
The HL-LHC projections are also given separately for comparison.

\section{Conclusions}

Under different scenarios, CLIC could take more data at the initial energy stage than assumed in 
the CLIC luminosity staging baseline.

Here, the effect on the Higgs coupling sensitivities of taking $4\,\abinv$ instead of $1\,\abinv$
at the initial stage has been presented, as an exercise to illustrate what could be obtained from
an alternative running scenario.  This could be achieved by running for 13 years instead
of 8 years, with an accelerator repetition rate of 100\,Hz instead of 50\,Hz.

\appendix

\section*{Acknowledgements}

This work is done in the context of the CLICdp Collaboration, and rescales results from many
detailed analyses that are presented in \cite{Abramowicz:2016zbo}.

This work benefitted from services provided by the ILC Virtual Organisation, supported by the national resource providers of the EGI Federation. This research was done using resources provided by the Open Science Grid, which is supported by the National Science Foundation and the U.S. Department of Energy's Office of Science.

\clearpage

\begin{minipage}{\linewidth}
  \begin{minipage}{0.495\textwidth}
    \begin{table}[H]
\begin{tabular}{lccc}
\toprule
Parameter & \multicolumn{3}{c}{Relative precision}\\
\midrule
& $350\,\GeV$ & + $1.4\,\TeV$ & + $3\,\TeV$\\
&$4\,\abinv$& + $2.5\,\abinv$& + $5\,\abinv$\\
\midrule
$\gHZZ$ & 0.3\,\% & 0.3\,\% & 0.3\,\% \\
$\gHWW$ & 0.5\,\% & 0.3\,\% & 0.3\,\% \\
$\gHbb$ & 1.0\,\% & 0.5\,\% & 0.4\,\% \\
$\gHcc$ & 2.2\,\% & 1.4\,\% & 1.1\,\% \\
$\gHTauTau$ & 1.5\,\% & 1.0\,\% & 0.8\,\% \\
$\gHMuMu$ & $-$ & 12.1\,\% & 5.6\,\% \\
$\gHtt$ & $-$ & 2.9\,\% & 2.9\,\% \\
\midrule
$g^\dagger_{\PH\Pg\Pg}$ & 1.3\,\% & 0.9\,\% & 0.7\,\% \\
$g^\dagger_{\PH\PGg\PGg}$ & $-$ & 4.8\,\% & 2.3\,\% \\
$g^\dagger_{\PH\PZ\PGg}$ & $-$ & 13.3\,\% & 6.6\,\% \\
\midrule
$\Gamma_{\PH}$ & 2.4\,\% & 1.5\,\% & 1.3\,\% \\
\bottomrule
      \end{tabular}
      \caption{ }\label{tab:MIResultsPolarised8020}
    \end{table}
  \end{minipage}
  \begin{minipage}{0.495\textwidth}
    \begin{figure}[H]
      \includegraphics[width=\linewidth]{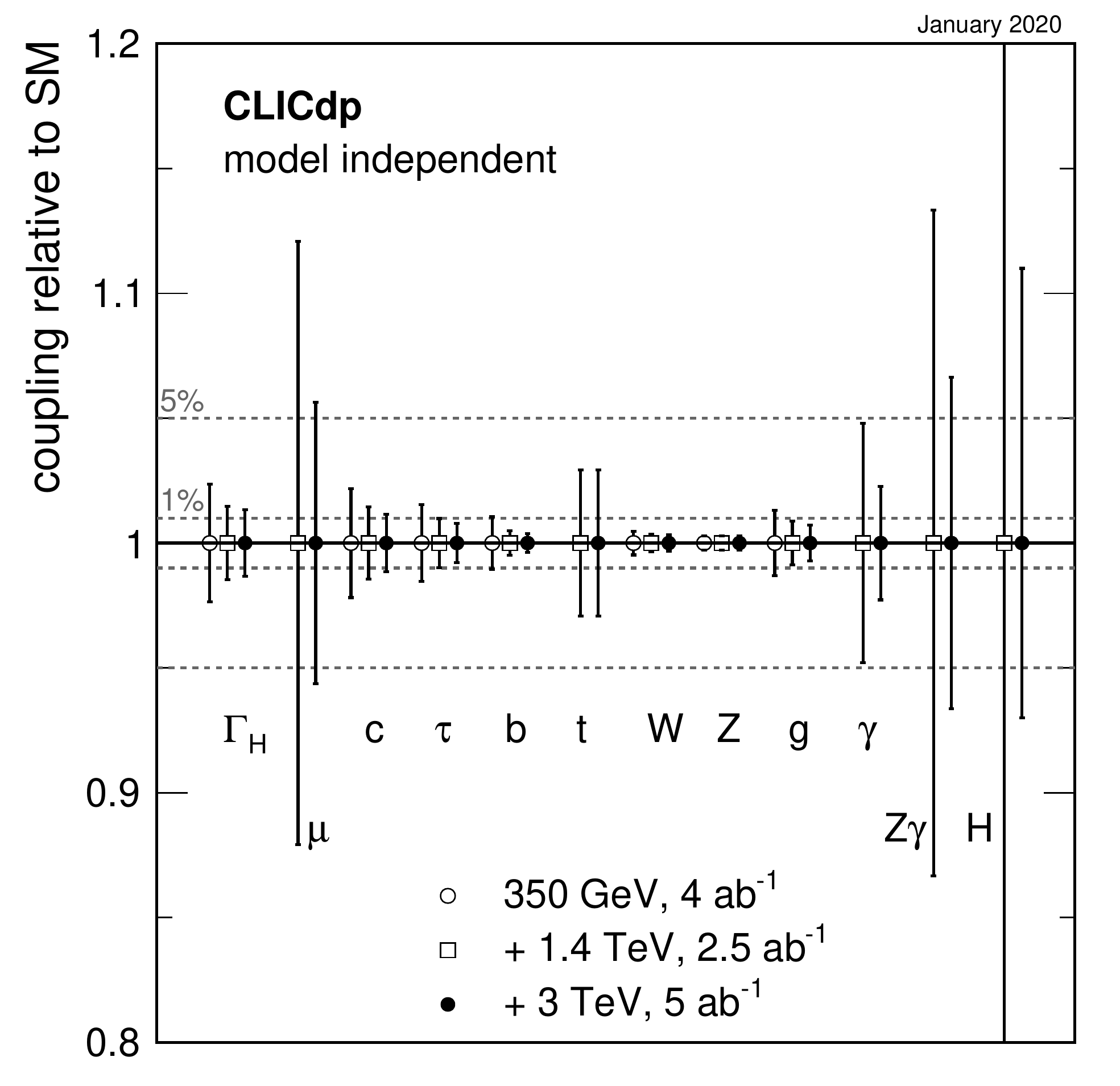}
      \caption{ }\label{fig:MIResultsPolarised8020}
    \end{figure}
  \end{minipage}
        Results of the model-independent fit assuming $4\,\abinv$ at $\sqrt{s}=350\,\GeV$, 
        for an illustrative scenario where more data is taken at the inital energy stage than in the CLIC baseline. 
        For $\gHtt$, the $3\,\TeV$ case has not yet been studied. The three
        effective couplings $g^\dagger_{\PH\Pg\Pg}$, 
        $g^\dagger_{\PH\PGg\PGg}$ and $g^\dagger_{\PH\PZ\PGg}$ are also included in the fit. 
        Operation with $-80\,\%$ ($+80\,\%$) electron beam polarisation is assumed for 
        $80\,\%$ ($20\,\%$) of the collected luminosity above 1\,\TeV, corresponding
        to the baseline scenario. 
\end{minipage}

\begin{minipage}{\linewidth}
\vspace*{1cm}
  \begin{minipage}{0.495\textwidth}
    \begin{table}[H]
\begin{tabular}{lccc}
\toprule
Parameter & \multicolumn{3}{c}{Relative precision}\\
\midrule
& $350\,\GeV$ & + $1.4\,\TeV$& + $3\,\TeV$\\
&$4\,\abinv$& + $2.5\,\abinv$& + $5\,\abinv$\\
\midrule
$\kappa_{\PH\PZ\PZ}$ & 0.2\,\% & 0.1\,\% & 0.1\,\% \\
$\kappa_{\PH\PW\PW}$ & 0.4\,\% & 0.1\,\% & 0.1\,\% \\
$\kappa_{\PH\PQb\PQb}$ & 0.6\,\% & 0.3\,\% & 0.2\,\% \\
$\kappa_{\PH\PQc\PQc}$ & 2.0\,\% & 1.4\,\% & 1.1\,\% \\
$\kappa_{\PH\PGt\PGt}$ & 1.4\,\% & 0.9\,\% & 0.7\,\% \\
$\kappa_{\PH\PGm\PGm}$ & $-$ & 12.1\,\% & 5.6\,\% \\
$\kappa_{\PH\PQt\PQt}$ & $-$ & 2.9\,\% & 2.9\,\% \\
$\kappa_{\PH\Pg\Pg}$ & 1.0\,\% & 0.8\,\% & 0.6\,\% \\
$\kappa_{\PH\PGg\PGg}$ & $-$ & 4.8\,\% & 2.3\,\% \\
$\kappa_{\PH\PZ\PGg}$ & $-$ & 13.3\,\% & 6.6\,\% \\
\bottomrule
\end{tabular}
\caption{ }\label{tab:MDResultsPolarised8020}
    \end{table}
  \end{minipage}
  \begin{minipage}{0.495\textwidth}
   \begin{figure}[H]
     \includegraphics[width=\linewidth]{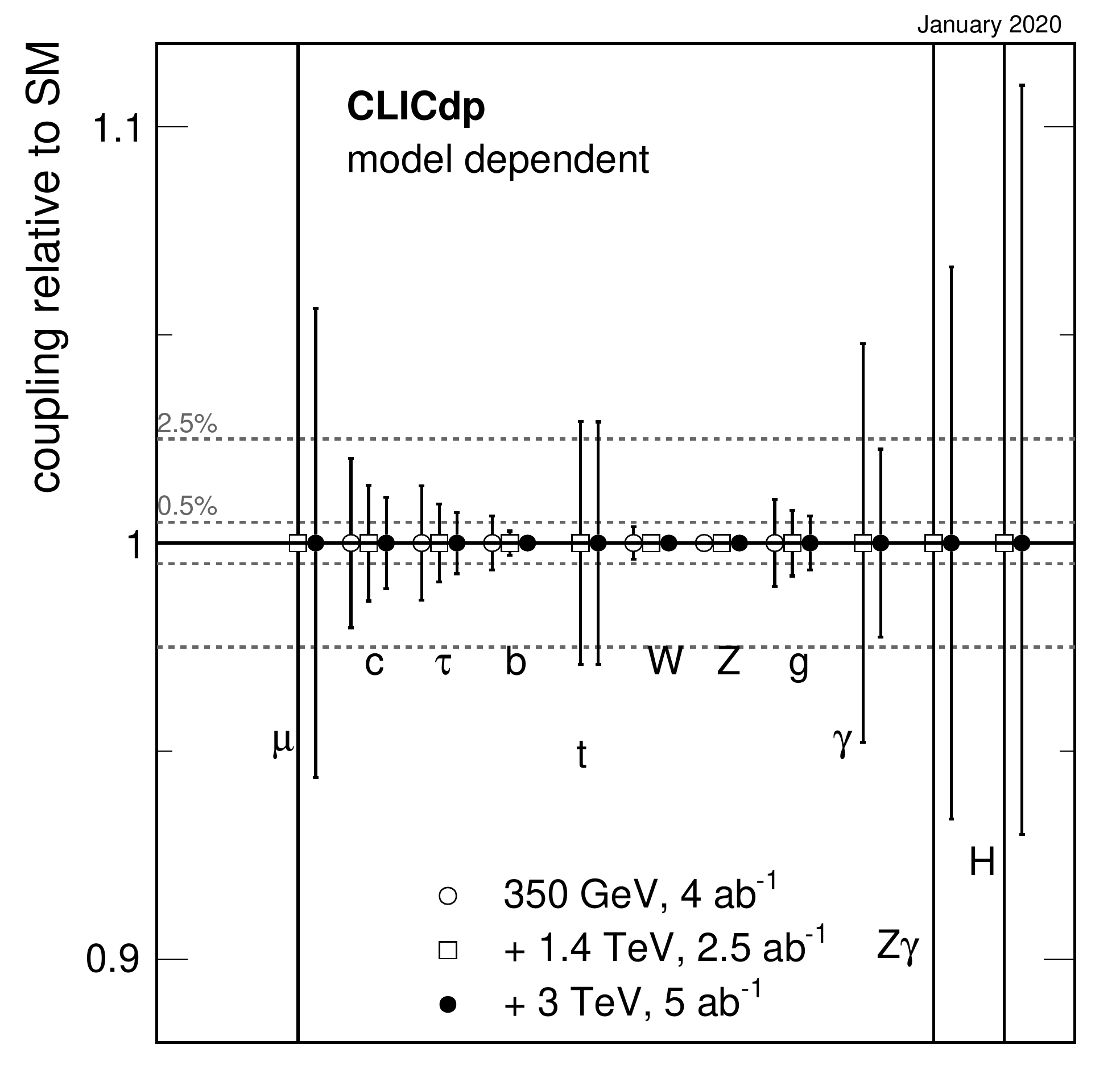}
     \caption{ }\label{fig:MDResultsPolarised8020}
   \end{figure}
  \end{minipage}
  Results of the model-dependent fit without theoretical uncertainties, assuming $4\,\abinv$ at $\sqrt{s}=350\,\GeV$, 
  for an illustrative scenario where more data is taken at the inital energy stage than in the CLIC baseline. 
  For $\kappa_{\PH\PQt\PQt}$, the $3\,\TeV$ case has not yet been
  studied. The uncertainty of the total width is calculated from the fit results. 
  Operation with $-80\,\%$ ($+80\,\%$) electron beam polarisation is assumed for 
  $80\,\%$ ($20\,\%$) of the collected luminosity above 1\,\TeV, corresponding to the baseline scenario. 
\end{minipage}

\begin{table}[t]
\centering
{ 
\begin{tabular}{ c | c | c | c c c }
\toprule
   &Benchmark   &HL-LHC   &\multicolumn{3}{c}{HL-LHC + CLIC} \\
   &    & & 380\,GeV   &  1.5\,TeV  & 3\,TeV  \\
 &    & & $4\,\abinv$& $2.5\,\abinv$&  $5\,\abinv$\\
\midrule
 \hspace{-0.2cm}$\gHZZ^{\mathrm{eff}}[\%]$& SMEFT$_{\textrm{ND}}$   & $3.6$   & $0.3$   & $0.2$   & $0.16$ \\
 \hspace{-0.2cm}$\gHWW^{\mathrm{eff}}[\%]$& SMEFT$_{\textrm{ND}}$   & $3.2$   & $0.3$   & $0.17$   & $0.14$ \\
 \hspace{-0.2cm}$g_{\PH\PGg\PGg}^{\mathrm{eff}}[\%]$& SMEFT$_{\textrm{ND}}$   & $3.6$   & $1.3$   & $1.3$   & $1.1$ \\
 \hspace{-0.2cm}$g_{\PH\PZ\PGg}^{\mathrm{eff}}[\%]$& SMEFT$_{\textrm{ND}}$   & $11.$   & $9.3$   & $3.2$   & $2.5$ \\
 \hspace{-0.2cm}$g_{\PH\Pg\Pg}^{\mathrm{eff}}[\%]$& SMEFT$_{\textrm{ND}}$   & $2.3$   & $0.9$   & $0.7$   & $0.60$ \\
 \hspace{-0.2cm}$\gHtt^{\mathrm{eff}}[\%]$& SMEFT$_{\textrm{ND}}$   & $3.5$   & $3.1$   & $2.1$   & $2.1$ \\
 \hspace{-0.2cm}$\gHcc^{\mathrm{eff}}[\%]$& SMEFT$_{\textrm{ND}}$   & $-$   & $2.1$   & $1.5$   & $1.2$ \\
 \hspace{-0.2cm}$\gHbb^{\mathrm{eff}}[\%]$& SMEFT$_{\textrm{ND}}$   & $5.3$   & $0.64$   & $0.42$   & $0.36$ \\
 \hspace{-0.2cm}$\gHTauTau^{\mathrm{eff}}[\%]$& SMEFT$_{\textrm{ND}}$   & $3.4$   & $1.0$   & $0.78$   & $0.65$ \\
 \hspace{-0.2cm}$\gHMuMu^{\mathrm{eff}}[\%]$& SMEFT$_{\textrm{ND}}$   & $5.5$   & $4.3$   & $4.1$   & $3.5$ \\
\midrule
 \hspace{-0.2cm}$\delta g_{1\PZ}[\times 10^{2}]$& SMEFT$_{\textrm{ND}}$   & $0.66$   & $0.027$   & $0.009$   & $0.007$ \\
 \hspace{-0.2cm}$\delta \kappa_{ \PGg}[\times 10^{2}]$& SMEFT$_{\textrm{ND}}$   & $3.2$   & $0.044$   & $0.023$   & $0.017$ \\
 \hspace{-0.2cm}$\lambda_{\PZ}[\times 10^{2}]$& SMEFT$_{\textrm{ND}}$   & $3.2$   & $0.022$   & $0.0051$   & $0.0018$ \\
\bottomrule
\end{tabular}
}
\caption{\label{tab:eft-global} 
Sensitivity at 68\% probability to deviations in the different effective Higgs couplings and anomalous triple gauge couplings from a Global SMEFT fit, using the benchmark SMEFT$_{\textrm{ND}}$ described in \cite{deBlas:2019rxi}. (The information about the other degrees of freedom included in the SMEFT$_{\textrm{ND}}$ fit in \cite{deBlas:2019rxi}, i.e. $g_{L,R}^f$, is omitted in this table.) These numbers can be compared to those of Table\,7 in \cite{deBlas:2019rxi}. For CLIC, results from the Z boson radiative return events are included.  Results are for an illustrative scenario where more data is taken at the inital stage, than in the CLIC baseline. 
}
\end{table}

\printbibliography[title=References]

\end{document}